\documentclass[twocolumn]{aastex701}

\newcommand{\be}{\begin{equation}}
\newcommand{\ee}{\end{equation}}
\newcommand{\bel}[1]{\begin{equation}\label{#1}}
\newcommand{\ba}{\begin{eqnarray}}
\newcommand{\ea}{\end{eqnarray}}
\newcommand{\bal}[1]{\begin{eqnarray}\label{#1}}

\newcommand{\Rtd}{R_{\rm td}}

\begin{document}

\title{Converged simulations of the nozzle shock in tidal disruption events} 

\author[0000-0001-7984-9477]{Fangyi (Fitz) Hu}
\affiliation{School of Physics and Astronomy, Monash University, Clayton VIC 3800, Australia}
\affiliation{OzGrav: The ARC Centre of Excellence for Gravitational Wave Discovery, Australia}
\email[show]{fangyi.hu@monash.edu}

\author[0000-0002-6134-8946]{Ilya Mandel}
\affiliation{School of Physics and Astronomy, Monash University, Clayton VIC 3800, Australia}
\affiliation{OzGrav: The ARC Centre of Excellence for Gravitational Wave Discovery, Australia}
\email{ilya.mandel@monash.edu}

\author[0000-0003-0856-679X]{Rebecca Nealon}
\affiliation{Centre for Exoplanets and Habitability, University of Warwick, Gibbet Hill Road, CV4 7AL Coventry, UK}
\affiliation{Department of Physics, University of Warwick, Gibbet Hill Road, CV4 7AL Coventry, UK}
\email{rebecca.nealon@warwick.ac.uk}

\author[0000-0002-4716-4235]{Daniel J. Price}
\affiliation{School of Physics and Astronomy, Monash University, Clayton VIC 3800, Australia}
\affiliation{Univ. Grenoble Alpes, CNRS, IPAG, 38000 Grenoble, France}
\email{daniel.price@monash.edu}



\begin{abstract}

When debris from a star that experienced a tidal disruption events (TDE) after passing too close to a massive black hole returns to pericenter on the second passage, it is compressed, leading to the formation of nozzle shocks (in the orbital plane) and pancake shocks (perpendicular to the orbital plane). Resolving these shocks is a long-standing problem in the hydrodynamic simulations of parabolic TDEs.  Excessive numerical energy dissipation or heating unrealistically expands the stream. In this Letter, we apply adaptive particle refinement to our 3D general relativistic smoothed particle simulations to locally increase the resolution near the pericenter.  We achieve resolutions equivalent to $6.55\times10^{11}$ particles, allowing us to converge on the true energy dissipation.  We conclude that only $4\times10^{-5}$ of the orbital energy is dissipated in nozzle shocks for a Sun-like star tidally disrupted by a $10^6$ solar-mass black hole, therefore the nozzle shocks are unlikely to be important in the evolution of TDEs.

\end{abstract}



\section{Introduction} \label{sec:intro}

Stars that pass sufficiently close to a massive black hole (MBH) are tidally disrupted, initiating a luminous transient: a tidal disruption event (TDE; \citealt{Rees:1988, Phinney:1989}).  The bound debris initially follows highly eccentric orbits with the closest approach distance equal to the pericenter of the initial orbit of the disrupted star.  

In order to release significant amounts of energy, some of this debris must experience large-angle deflections.  These would allow some of the gas to move onto orbits that take it closer to the MBH than the initial pericenter distance, converting gravitational potential energy into kinetic energy.  Once there is sufficient gas on trajectories with large deflection angles, aided by a cascade of secondary and tertiary collisions and scatterings closer to the MBH, an elliptical disk or torus will form \citep{Bonnerot:2016,Hayasaki:2016,Hu2024,SteinbergStone:2024}. Subsequent interactions within that disk and between the disk and returning streams will continue to  cancel angular momentum and release energy, with additional efficient energy release occurring once accretion onto the MBH starts. This prompt energy release could explain the relatively fast rise of TDE lightcurves.
For example, when fitting TDE light curves with a model in which the luminosity was proportional to the mass return rate while allowing for a delay, \citet{Mockler:2019} found that the requisite time delay was consistent with zero.

What is the physical process responsible for the initial large-angle deflections?  Several mechanisms have been proposed.  Nozzle shocks or vertical pancake shocks near the pericenter are one popular mechanism \citep[e.g.][]{Brassart2008, Stone2013, Coughlin2016, Guillochon2009, Coughlin2020a}.  These have been argued to dissipate sufficient energy (a few to tens of percent of the orbital energy) to allow for large-angle deflections, with subsequent collisions near the MBH and/or accretion promptly taking over as the dominant energy source \citep[e.g.,][]{SteinbergStone:2024}.   Other authors have advocated for collisions between debris streams, sometimes referred to as `apocenter' or `outer shocks', as the primary mechanism for energy dissipation and large-angle deflections \citep[e.g.,][]{Piran:2015, LuBonnerot:2020}. 

The `stream fanning' associated with the nozzle shock seen in 3D simulations of stars on parabolic orbits \citep{Guillochon:2014,Ryu:2023, Price2024a, SteinbergStone:2024} is in contrast to simulations of stars on bound orbits which did not find significant energy dissipation associated with the nozzle shock \citep{Hayasaki:2013,Bonnerot:2016,Hayasaki:2016,Hu2024}. Furthermore, simple analytical considerations (Section~\ref{sec:ana}; see also \citealt{Guillochon:2014}) and 2D local models \citep{BonnerotLu:2022} suggest that nozzle shocks cannot be as efficient as implied by 3D global simulations. The observed levels of energy dissipation in nozzle shocks in parabolic TDE simulations are likely overestimated due to the need to resolve the extremely short length scale (estimated at $H\sim 10^{-3}R_*$ in the 2-d simulations of \citealt{BonnerotLu:2022}) associated with the stream compression \citep[see resolution study in][]{Price2024a}.  

We carry out a smoothed particle hydrodynamics (SPH) simulation of a Sun-like star being disrupted near a $10^6 M_{\odot}$ MBH on a parabolic orbit, and apply adaptive particle refinement (APR) \citep{Nealon2025} to locally increase the resolution at the pericenter for a resolution study.
We have achieved converged measurements of nozzle shock dissipation energy at a resolution equivalent the $4.1\times10^{10}$ particles.  This study unambiguously shows that the fraction of energy dissipated in a nozzle shock is $\lesssim 10^{-4}$ for the chosen parameters, so the nozzle shock cannot be responsible for large deflection angles. 

\section{Analytical considerations} \label{sec:ana}

In this section, we outline an analytical explanation for why nozzle shocks and pancake shocks are not sufficient to drive rapid energy dissipation and stream circularisation.  These order-of-magnitude calculations are supported by a detailed computational model in the following section.

A star of mass $m$ and radius $r$ will be tidally disrupted by a MBH of size $M$ at the tidal disruption radius $\Rtd$ at which the tidal force from the massive body exceeds the object's gravity:
\ba
\Rtd &\sim& \left(\frac{M}{m}\right)^{1/3} r \\
&\approx& 0.5\, \textrm{au}\left(\frac{M}{10^6 M_\odot}\right)^{1/3} \left(\frac{m}{M_\odot}\right)^{-1/3} \left(\frac{r}{R_\odot}\right) . \nonumber
\ea
Suppose that the star approaches the MBH on a parabolic orbit with pericenter $R_\mathrm{p} = \Rtd$ (i.e., $\beta = 1$).  The most bound debris from the disrupted star will follow a trajectory with semi-major axis $a_\mathrm{mb}$ given by 
\ba
a_\mathrm{mb} &\sim& \frac{\Rtd^2}{r} \sim \left(\frac{M}{m}\right)^{1/3} \Rtd \\
&\approx& 50 \left(\frac{M}{10^6 M_\odot}\right)^{2/3} \left(\frac{m}{M_\odot}\right)^{-2/3} \left(\frac{r}{R_\odot}\right)\, \textrm{au}. \nonumber
\ea
Since this most-bound debris has a pericenter at $\Rtd$ and an apoapsis at $\approx 2 a_\mathrm{mb}$, it will have an eccentricity of order 
\ba
1-e_\mathrm{mb} &=& \frac{\Rtd}{a_\mathrm{mb}} \sim \left(\frac{m}{M}\right)^{1/3} \\
&\approx& 0.01 \left(\frac{M}{10^6 M_\odot}\right)^{-1/3} \left(\frac{m}{M_\odot}\right)^{1/3}, \nonumber
\ea
i.e., $e_\mathrm{mb} \approx 0.99$ for our fiducial values.

As has been widely pointed out in the literature \citep[e.g.,][]{Piran:2015}, accretion is impossible from such an eccentric flow whose pericenter is much larger than the Schwarzschild radius of the MBH, 
\be
\Rtd \gg R_s = \frac{2 G M}{c^2} \approx 0.02 \frac{M}{10^6 M_\odot}\, \mathrm{au},
\ee
and energy must be dissipated before the debris can circularise and accretion can commence.


The bulk velocity of the flow at pericenter is of order 
\ba
v_\mathrm{bulk} &\approx& \sqrt{\frac{2G M}{\Rtd}} = \sqrt{2} \left(\frac{M}{m}\right)^{1/3} \sqrt{\frac{G m}{r}} \\
&\equiv& \left(\frac{M}{m}\right)^{1/3} v_\mathrm{th}, \nonumber
\ea
where we denote the virial-theorem order-of-magnitude thermal velocity of the incoming star, and hence stream, by $v_\mathrm{th}$.   The Mach number of the stream \citep{Kochanek:1994} is then 
\be
\mathcal{M} = \left(\frac{M}{m}\right)^{1/3}.
\ee
For $M = 10^6 m$, the bulk velocity is two orders of magnitude greater than the thermal velocity and the Mach number is $\sim 100$.  That means the amount of energy dissipated during a nozzle passage is of order $dE/E \sim v_\mathrm{th}^2 / v_\mathrm{bulk}^2 \sim 10^{-4}$ -- very unlikely to be sufficient for rapid circularisation.  

The same calculation applies to the pancake shock from vertical motion.  The largest vertical velocity particles can have at pericenter can be obtained by considering the inclined ballistic trajectories of disrupted debris.  Since initially disrupted gas is at most a distance $r$ above or below the equatorial plane while at a distance $\Rtd$ from the star, the orbits are inclined by at most $r/\Rtd$, so the maximum vertical velocity at pericenter is $v_{z} \approx (r/\Rtd)v_{\rm bulk}$.  Again, the ratio of the bulk velocity to the vertical velocity is $v_\mathrm{bulk}/v_\mathrm{z} \sim (M/m)^{1/3} \approx 100$, and the energy dissipation due to pancaking is a quantity of order $dE/E \sim 10^{-4}$. The same conclusion was reached by previous authors. In particular \citet{Guillochon:2014} estimated $dE/E \sim (M/m)^{-2/3}$ for the nozzle shock dissipation, also giving $10^{-4}$ for $\beta = 1$. 

Recombination of adiabatically cooling debris at large distances from the MBH can inject energy into the stream and has been proposed as another source of energy dissipation \citep{SteinbergStone:2024}.  The total energy of one proton following the most bound trajectory is
\ba
|E_\mathrm{mb}| &=& \frac{G M m_p}{2 a_\mathrm{mb}} \\
&\approx& 9 \times 10^4 \left(\frac{M}{10^6 M_\odot}\right)^{1/3} \left(\frac{m}{M_\odot}\right)^{2/3} \left(\frac{r}{R_\odot}\right)^{-1}\,  \textrm{eV}. \nonumber
\ea
This is 4 orders of magnitude greater than the $13.6$ eV released by hydrogen recombination.  Thus, once again, $dE/E \sim 10^{-4}$, so recombination leaves $v_\mathrm{th} \sim v_z \sim 0.01 v_\mathrm{bulk}$.  (Note that we refer to the relative kinetic energy between debris components as $dE$, not to the spread of kinetic energies in the MBH frame, which as \citet{Ryu:2023,SteinbergStone:2024} rightly point out is of order $\delta E \sim v_\mathrm{th}  v_\mathrm{bulk}$.) One caveat is that even though recombination energy injection itself is insignificant, it could lead to a vertical expansion of the stream when recombination energy is injected far from the MBH.  \citet{SteinbergStone:2024} estimate an expansion by a factor of 3, which would increase the vertical velocity by the same factor, and thus potentially increase the fraction of energy dissipated in a pancake shock to 0.1\%; the simulations described in the next section do not include recombination energy injection.

Perhaps a more intuitive way to think about this is in terms of stream spreading or flaring rather than just energy dissipation.  With perpendicular motion suppressed by two orders of magnitude relative to bulk forward motion, the stream can be thought of as cars driving on a highway at 100 km s$^{-1}$ that may slowly drift sideways at 1 km s$^{-1}$.  While there will certainly be some damage during such collisions, we do not expect post-collision trajectories to diverge by more than $1/\mathcal{M} \sim 1/100$.  Stream broadening by this small fraction is not sufficient to deflect matter onto trajectories that would allow it to closely approach the MBH.

This leaves stream-stream collisions driven by relativistic pericenter precession as the only viable option for rapid energy dissipation and change in trajectory by large deflection angles.  
There are two possible collision locations to consider.  

First, consider collisions near pericenter itself.  Although streams of more tightly bound particles on the second post-disruption pericenter passage pass through pericenter at a very similar location to the less bound particles on their first post-disruption passage, the streams are angled because of pericenter precession, providing a perpendicular component of the velocity $v_\mathrm{p} \approx \theta_\mathrm{p} v_\mathrm{bulk}$.  Here, $\theta_\mathrm{p} \sim 3\pi R_\mathrm{g} / R_\mathrm{p}$ is the pericenter precession angle per nearly parabolic orbit, where the gravitational radius is $R_\mathrm{g} = GM / c^2$.  Assuming, as we have done, that the pericenter radius is the tidal disruption radius, $R_\mathrm{p} = \Rtd$, the amount of precession per orbit is 
\be 
\theta_\mathrm{p} \approx 0.2  \left(\frac{M}{10^6 M_\odot}\right)^{2/3} \left(\frac{m}{M_\odot}\right)^{1/3} \left(\frac{r}{R_\odot}\right)^{-1}\, .
\ee
The fraction of bulk kinetic energy released is thus of order 4\% for the fiducial parameters we have chosen.  Meanwhile, the degree of flaring of the stream after the collision is $\sim 0.2$ radians $\approx 11^\circ$ for the chosen parameters.  This collision at pericenter requires at least the most bound debris to complete two orbits, in order to catch up and collide with less strongly bound debris completing its first post-disruption orbit.

The second type of crossing is between the precessed stream of more tightly bound particles emerging from a pericenter passage and the incoming stream of less tightly bound particles \citep{Shiokawa:2015, Andalman:2022}.  This collision happens further from the MBH, so at lower velocities, but at a much steeper angle.  In fact, \citet{Dai:2015} showed that the typical angles range from 50 to above 150 degrees, i.e., an order unity fraction of the bulk kinetic energy is dissipated in such distant stream-stream collisions.  Correspondingly, the flaring angle is also of order unity: the debris is emitted over a very broad angle from the point of collision, as argued by past studies \citep[e.g][]{LuBonnerot:2020, Hu2024, SteinbergStone:2024}.  Importantly, this crossing first happens at between 1 and 1.5 times the orbital period of the most bound debris following the initial disruption, i.e., sooner than the precession-induced collisions at pericenter.  Thus, it makes the collision at pericenter irrelevant.  



Whether the stream-stream collisions leading to the formation of an elliptical disk are themselves sufficient to explain the full observed luminosity \citep{Piran:2015} may depend on the amount of precession relative to the opening angle of the most-bound ellipse.  Streams will cross within a distance $a_\mathrm{mb}$ of the MBH if $\theta_\mathrm{p} \gtrsim \sqrt{1-e^2}$, where the latter is the ratio of the minor and major axis of the ellipse followed by the most bound debris.  For our fiducial parameters, the two quantities are comparable.  The bulk stream velocity is $\gtrsim \sqrt{GM/a}$ for collisions that happen within the ellipse semi-major axis $a$ of the MBH; for our fiducial parameters, that is $\gtrsim 0.01 c$, so the energy release rate is $\gtrsim 10^{-4} \dot{M} c^2$.  On the other hand, if the precession angle per orbit is less than $\sqrt{1-e^2}$, stream-stream collisions will only happen near apoapsis.  The bulk velocity near apoapsis is $\sim \sqrt{GM/a} \sqrt{(1-e)/(1+e)} \approx 0.0007 c$ for our fiducial parameters, leading to an energy release rate of only $\lesssim 10^{-6} \dot{M} c^2$ from collisions there, likely too little to directly explain the observed luminosities, however still able to deflect gas by a large angle and lead to follow-up collisions or accretions.

\begin{figure}
    \centering
    \includegraphics[width=\linewidth]{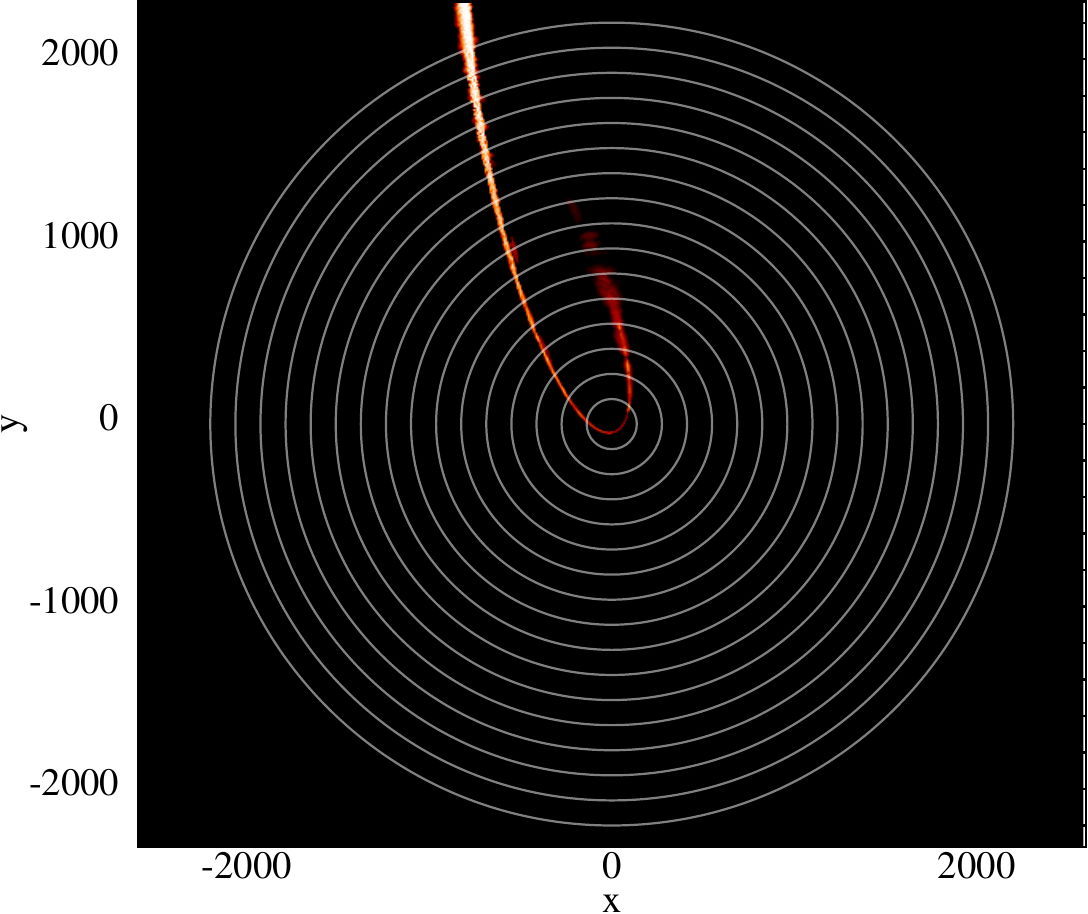}
    \caption{Demonstration of the APR boundaries used (16 levels in the figure). The spherical boundaries are centred on the MBH. The outermost boundary is at $r=2200 r_{\rm g} \approx 3.25\times10^{14} {\rm cm}$ with $\Delta r = 137.5 r_{\rm g} \approx 2.03\times 10^{13} {\rm cm}$ between each boundary, where $r_{\rm g} = GM/c^2$ is the gravitational radius of the MBH. $r=2200r_{\rm g}$ is the same in all simulations whereas $\Delta r = 1100, 550, 275, 183.3$ for 2, 4, 8, 12 level simulations respectively.}
    \label{fig:split_demo}
\end{figure}

\begin{figure*}
    \centering
    \includegraphics[width=\linewidth]{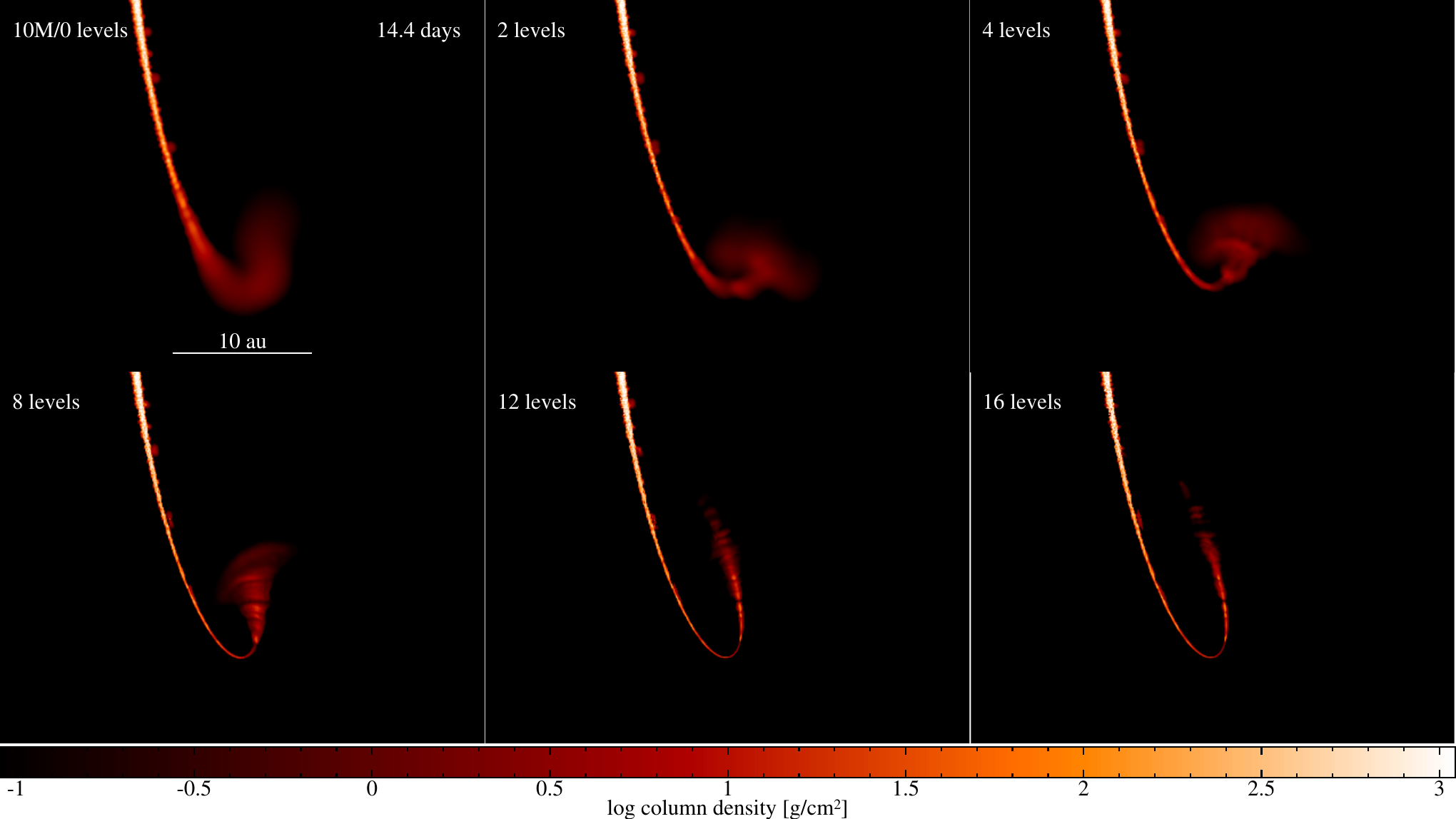}
    \caption{Snapshots of the simulations with 0, 2, 4, 8, 12, 16 levels of splitting at $t=14.4$ days after stellar disruption (first pericenter passage). The eccentricity of the most bound gas ($e_{\rm mb,simulated} \sim 0.96$) is smaller than predicted in Section~\ref{sec:ana} ($e_{\rm mb,analytic} \sim 0.99$) due to the heating from nozzle and pancake shocks on the first pericenter passage.
    The amount of energy dissipation at pericenter and thus stream fanning post-pericenter decreases with resolution until it converges at 12 levels of refinement (equivalent to a global 41B particle simulation). The binary simulation dumps plotted are available on Zenodo: 10.5281/zenodo.17225834.}
    \label{fig:split}
\end{figure*}

\section{Nozzle shock} \label{sec:nozzle}
Previous SPH studies \citep[e.g.][]{Price2024a} performed 3D general relativistic simulations of parabolic TDEs with typical highest resolutions of $\lesssim10^7$ particles. They found that the nozzle shock is highly resolution dependent, i.e., the dissipated heat and the amount of stream flaring or deflection is reduced with increasing resolution.  Their simulations were not converged. 

\subsection{Adaptive particle refinement} \label{sec:apr}
To solve the resolution problem at the pericenter without an intractable computational time, we use the newly developed adaptive particle refinement (APR) in {\sc phantom} \citep{Price2018,Liptai2019a, Nealon2025}. The APR implementation allows for different resolution levels to occur in the same simulation and we choose the highest resolution region to be centred on the black hole. As individual particles approach the black hole they are split into two at each APR refinement boundary to increase the resolution, and after pericenter passage as they move away from the black hole two particles are merged into one to return to the original resolution. In our subsequent simulations we use multiple levels of refinement such that the particles are split and merged numerous times.

\subsubsection{Splitting along a geodesic} \label{sec: split geo}
\citet{Nealon2025} found that it was beneficial to split particles tangential to the resolution boundaries so that newly created high resolution particles didn't accidentally land in the lower resolution region which the parent particle just exited (see their Figure 1). For a TDE with the high resolution region centred on the black hole, this means that particles would be split perpendicular to their geodesic which then changes their orbits.

We modified the particle splitting to avoid widening the debris stream and conserve the particle orbit and energy since these are critical in the case of TDEs.  Here, instead of splitting tangential to the resolution boundary, particles are split along their individual geodesic. When a particle has been identified to be split, we evolve the particle forward and backward in time for $\Delta t$ along its geodesic for a distance of $r_{\rm sep} = 0.2 h_{\rm parent}$, where $h_{\rm parent}$ is the smoothing length of the parent particle, i.e. $\Delta t =  r_{\rm sep} / | v|_{\rm parent}$. We integrate the momentum $\vec{p}$ and position $\vec{x}$ of the particle with only the external force due to the background metric (method 5 in Appendix~\ref{app:benchmark}). 
We complete the integration with at least 10 steps to ensure an accurate geodesic. The children particles will be placed roughly $\pm r_{\rm sep}$ from the parent particle along the geodesic. Appendix~\ref{app:benchmark} details a comparison with a few different methods of splitting along a geodesic, demonstrating that the above method is the most accurate for a TDE among those considered. 

With our updated implementation, it is possible that one of the children particles may be placed back across the refinement boundary into the pre-splitting region which the parent just left. \citet{Nealon2025} identified this as a possible issue; however, we did not find this to be the case for our application here.

\subsubsection{Relaxation}

We found that occasionally after splitting, children particles form additional shocks with nearby particles that were absent in simulations without APR - this is likely due to the increased noise at the resolution boundaries due to the split/merge procedure identified by \citet{Nealon2025} despite checking that newly split particles are sufficiently far from existing particles, as in \citet{Franchini:2022nh}. Instead we found that particles could be sufficiently far but their velocities meant that in the next timestep they would create a shock. In the case of the TDE stream, these shocks release excessive heat which expands or even disrupts the debris stream.

To combat this, we relax the system after each split or merge with shuffling as described in Section 1.4 of \citet{Nealon2025}. \citet{Nealon2025} found that this relaxing procedure gradually and monotonically decreased the noise for splitting events. In our simulations, however, this was not the case and so instead of stopping relaxation based on the kinetic energy metric or upon reaching 50 shuffles of each particle to be relaxed, we make only two shuffles each time. Each shuffle has a timestep of $\Delta t=0.1h/c_{\rm s}$, where $h$ and $c_{\rm s}$ are the smoothing length and sound speed of each particle respectively. The first shuffle resolves the shocks forming due to particle placement when splitting (if any) and the second shuffle allows for additional smoothing of the gas. We also restrict each shuffle to have $|dx| \leq h$ \citep[as in][]{Nealon2025} to prevent overexpansion of the debris stream.  The thermal energy is fixed during the relaxation process. 

\subsection{Resolution study}
To explore the resolution of a nozzle shock, we use our $10^7$-particle TDE simulation of a $1R_{\odot}$, $1M_{\odot}$ main-sequence star around a $M_\mathrm{MBH} =10^6 M_{\odot}$ MBH on a parabolic orbit with $\beta=1$ and perform 5 additional simulations with APR. We performed the simulations using general relativistic SPH \citep{Liptai2019a} with a fixed background Schwarzschild metric for the black hole. We use the same MESA star \citep{Paxton2011, Paxton2013, Paxton2015, Paxton2018, Paxton2019} as in \citet{Hu2024}. The star was relaxed into hydrostatic equilibrium using the procedure described in Appendix~C of \citet{Lau:2022} before placing it on a parabolic orbit at an initial separation of $3000r_{\rm g}\approx 4.4\times10^{14}$ cm from the black hole. Note that the chosen parameters lead to a partial tidal disruption, with the core of the star surviving; however, this does not affect the efficiency of the nozzle shock at dissipating energy of the leading edge of the stream (the most bound material), which is the focus of our study.

The APR simulations have 2, 4, 8, 12, or 16 levels of refinement, i.e., are equivalent to 40M, 160M, 2.56B, 41B, 655B particles globally in a simulation at the highest resolution near the MBH. 

We set all the APR regions to be spherical, centering on the MBH at the origin of our simulation (see Figure~\ref{fig:split_demo}). The outermost resolution boundary in all 6 simulations is at a radius of $2200\ r_{\rm g} \equiv 2200 G M_\mathrm{MBH} c^{-2} = 3.25\times 10^{14}$ cm. The rest of the boundaries are spaced evenly between the outermost boundary and the origin (see circles in Figure~\ref{fig:split_demo}). We note that the distance between each resolution region allows for $\gtrsim 100$ sound crossings of the gas, allowing particles to fully relax between splitting and merging events and reducing the total noise \citep{Nealon2025}.

Our results are shown in Figure~\ref{fig:split} for the different resolutions. The resolution dependence is most apparent for the gas that has just passed through pericentre, with the lower resolution simulations showing noticeable spreading of the orbital streams, indicating more energy is dissipated in the nozzle and pancake shocks. The width of the stream is consistent between the 12 and 16 refinement level simulations, indicating convergence.

We note the orbital period of the most bound gas in Figure~\ref{fig:split} is 14.4 days, which corresponds to $e_{\rm mb}=0.96$ with our choice of pericenter distance ($R_{\rm p} = R_{\rm td}=100 R_\odot=6.96\times10^{12}$ cm), not 0.99 as predicted in Section~\ref{sec:ana}. This is due to the heating from nozzle and pancake shocks within the star during the first pericenter passage spreads gas onto a wider range of orbits.





\begin{figure}
    \centering
    \includegraphics[width=\linewidth]{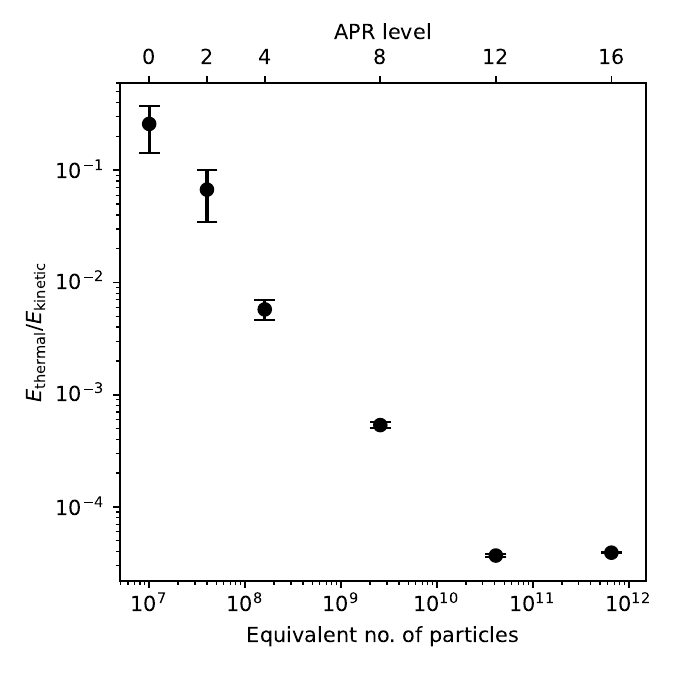}
    \caption{The mean fraction of energy dissipated near the pericenter. The fraction drops almost linearly in log scales until convergence at 12 levels of splitting. Using linear fits to the four points before convergence and separately to the two points after convergence, we estimate the nozzle shock converge at approximately $2.24\times10^{10}$ particles. The error bars represent the statistical standard errors of our measurements and not the true uncertainties of the energy fractions. Data is available on Zenodo: 10.5281/zenodo.17225834.}
    \label{fig:circularisation efficiency}
\end{figure}

\subsubsection{Energy dissipation} \label{sec:circ eff}
We represent the fraction of energy dissipated near the pericenter as the ratio of thermal energy to kinetic energy, following the descriptions in Section~\ref{sec:ana}. 

To find the thermal energy and kinetic energy of the stream at pericenter, we first divide the plane (Figure~\ref{fig:split}) into angular beams with opening angles of $\pi/60=3^\circ$ as viewed from the MBH. We then calculate the location of pericenter as the angular beam with the local maximum of the kinetic energy from our highest resolution, with an additional beam on either side to account for the uncertainties. We average the properties of the $n_\mathrm{p}$ particles within each beam and use the corresponding standard deviation $\sigma$ as our measure of the uncertainty in the specific thermal and kinetic energies (more details are available in Appendix~\ref{app:energy_appendix}).

In Figure~\ref{fig:circularisation efficiency}, we plot the energy fraction at the approximate pericenter against the equivalent number of particles or APR levels. The error bars represent the standard errors, i.e. $\sigma/\sqrt{n_{\rm p}}$. 

The energy fractions dissipated with 0 and 2 APR levels (1 and 4$\times10^7$ particles respectively) are $\gtrsim0.1$, which is similar to \citet{SteinbergStone:2024} who used $2\times10^7$ grid cells. The fraction of dissipated orbital energy drops with resolution and converges to $\sim 4\times10^{-5}$ at 12 levels. Using a linear fit to the four points pre-convergence (0, 2, 4, 8 levels) and two points post-convergence (12, 16 levels), we find the nozzle shock would converge at approximately $2\times10^{10}$ particles. While a considerable number of particles, this is not nearly as pessimistic as estimated by \citet[][their Equation~11]{BonnerotLu:2022} who predicted that $\approx 10^{14}$ SPH particles would be required.

\subsubsection{Implications}
Following Section~\ref{sec:ana}, an energy dissipation fraction of $4\times10^{-5}$ corresponds to a Mach number of 160, which gives a deflection angle of 0.006 rad $=0.34^\circ$. This is too small to significantly affect the subsequent evolution of the debris stream. The same conclusion was previously reached by \citet{BonnerotLu:2022} based on local 2D simulations of the nozzle shock. We expect that stream-stream collisions are instead responsible for large-angle deflections, allowing debris to reach closer to the MBH and leading to the formation of an eccentric disk, as indeed found at higher resolution in simulations, albeit without being fully converged \citep{Price2024a}.

However, the timescale on which the MBH is obscured by a reprocessing cloud of gas after the second pericenter passage of the most bound debris, with large optical depths along all lines of sight preventing X-rays generated close to the MBH from reaching a distant observer, may be overestimated in 3D parabolic TDE simulations due to the overestimated lateral expansion of the stream. Mitigating this is that prompt smothering is also found in TDE simulations of stars on bound orbits, where there are no known resolution issues \citep{Hu2024}.


If the MBH were not as promptly obscured, it could allow X-rays generated by shocks or accretion close to the MBH to escape, at least along some lines of sight \citep{Dai:2018}.  This could naturally explain prompt X-rays observed from some TDEs \citep[e.g.][]{Auchettl2017}. The details require future simulations to study.





\section{Conclusion} \label{sec:conclusion}
We have utilised the newly developed adaptive particle refinement method \citep{Nealon2025} to locally increase the resolution of simulations at the pericenter to study the nozzle shock. The simulation converges at 12 levels of splitting, i.e. equivalent to $4.1\times10^{10}$ particles globally. The fraction of orbital energy dissipated during nozzle shock converges to $\sim 4\times10^{-5}$ which also agrees with analytical predictions within a factor of a few. This corresponds to a stream expansion of $\sim 0.006$ rad, therefore unlikely to be critical for altering the orbits of the debris stream as well as releasing the first light from TDEs.

\begin{acknowledgments}
We acknowledge the CPU time on OzSTAR and Ngarrgu Tindebeek funded by the Victorian and Australian governments and Swinburne University. F.H.~and I.M.~acknowledge support from the Australian Research Council Centre of Excellence for Gravitational Wave Discovery (OzGrav) through project No. CE230100016.  I.M.~thanks Elena Rossi and Nick Stone for illuminating discussions.  D.P.~thanks Cl\'ement Bonnerot for helpful discussions. R.N.~acknowledges the hospitality of Monash University.
\end{acknowledgments}

%





\appendix

\section{Alternative particle splitting methods} \label{app:benchmark}
The splitting method introduced in \citet{Nealon2025} conserves mass, angular momentum and kinetic energy because the newly split particles inherit the velocity of the original particle. These particles are placed along a vector that is tangential to the refinement boundary at a distance of $r_{\rm sep} = 0.2 h_{\rm parent}$, the smoothing length of their parent particle. In the implementation used, the position of particles that are placed too close to a nearby particle is adjusted following \citet{Franchini:2022nh}.

However, in the case of a tidal disruption event around an MBH, the refinement boundary is perpendicular to the streamlines of the disrupted star. When particles cross the refinement boundary and are then split with the above implementation, the children particles are placed onto a different geodesic and hence a different orbit and energy. Changing the orbit of newly refined particles could problematically change their fate, e.g. changing a particle that would have been accreted to being ejected. In addition to this, spliting perpendicular to the refinement boundary does not create a continuous stream while passing the pericenter which is not ideal. To mitigate this problem we have tested five new ways of calculating the children particle's positions and velocities with the goal of conserving their orbit and energy. We use $r_{\rm sep} = 0.2h_{\rm parent}$ for all of the following methods:
\begin{enumerate}
    \item As in \citet{Nealon2025}: particles are split tangential to the APR region boundary with children placed $r_{\rm sep}$ from the parent particle. Children inherit all the other properties of the parent (velocity, thermal energy, etc.) and their smoothing length is scaled with their new mass to conserve density.
    \item Split along the velocity vector: children inherit the properties of their parents but instead of being placed perpendicular to the refinement boundary, they are placed a distance $r_{\rm sep}$ along their individual velocity unit vector $\tilde{v}_{\rm parent}$.
        \begin{equation}
            \vec{x}_{\rm{child}} = \vec{x}_{\rm{parent}} \pm r_{\rm{sep}}\tilde{v}_{\rm{parent}}
        \end{equation}
    \item Along a geodesic with the initial metric: particle positions are extrapolated along the particle trajectory, assuming the initial metric. We use $\Delta t = r_{\rm sep}/|v|_{\rm parent}$ and a simple first order prediction given by
     \begin{equation} \label{eq:flip}
                \vec{p}_{\rm parent} \Rightarrow \vec{p}_{\rm child, front}; -\vec{p}_{\rm parent} \Rightarrow \vec{p}_{\rm child, behind}
        \end{equation}
        \begin{equation}
            \textrm{for each child: front \& behind }
            \left\{ 
            \begin{array}{l}
                \ \ \vec{p}_{\rm child} = \vec{p}_{\rm child} + \vec{f}^{\rm ext}_{\rm parent}(\vec{x}_{\rm parent}) \Delta t \\
                \ \ \vec{p}_{\rm child} \rightarrow \vec{v}_{\rm child}, \\
                \ \ \vec{x}_{\rm child} = \vec{x}_{\rm parent} + \vec{v}_{\rm child}\Delta t,
                \end{array}\right .
        \end{equation}
         \begin{equation} \label{eq:flip back}
            \begin{array}{l}
                \vec{v}_{\rm child, front} \Rightarrow \vec{v}_{\rm child, front}; - \vec{v}_{\rm child, behind} \Rightarrow \vec{v}_{\rm child, behind} \\
                \vec{p}_{\rm child, front} \Rightarrow \vec{p}_{\rm child, front}; - \vec{p}_{\rm child, behind} \Rightarrow \vec{p}_{\rm child, behind}
            \end{array}
        \end{equation}
    where $\vec{p}$ is the conserved momentum, $\vec{f}^{\rm ext}$ is the external force from the background metric, and `$\rightarrow$' represents solving for primitive variables ($\vec{v}$) from the conserved variables ($\vec{p}$) (see \citet{Liptai2019a} for more details). The momentum $\vec{p}$ of the child that evolves backwards in time, i.e. behind the parent, is flipped in direction to evolve backwards in time (Equation~\ref{eq:flip}) and flipped back after calculation (Equation~\ref{eq:flip back}).
    \item Along a geodesic but updating velocity: here we update the children's velocity to $\vec{v}_{n+1}$ according to the initial metric while extrapolating the particle's trajectory to $\vec{x}_{n+1}$. This integration is completed with $N\geq10$ timesteps until $\Delta t = r_{\rm sep}/|v|_{\rm parent}$ for accuracy, with $\delta t = \min \left(0.1 u_t, 0.1\Delta t, \Delta t_{\rm ext}\right)$, where $u_t$ is the code time unit and $\Delta t_{\rm ext}$ is the external force time step \citep[see][]{Liptai2019a}. 
    \begin{equation} \label{eq:flip 2}
                \vec{p}_{\rm parent} \Rightarrow \vec{p}_{\rm 0, child, front}; -\vec{p}_{\rm parent} \Rightarrow \vec{p}_{0, \rm child, back}
    \end{equation}
    \begin{equation}
        \begin{array}{cc}
            \textrm{for each child: front \& behind } \\
            \textrm{loop over } n \textrm{ until } \Delta t \textrm{ is reached} 
        \end{array}
        \left\{
        \begin{array}{l}
                \vec{p}_{n+1} = \vec{p}_{n} + \vec{f}^{\rm ext}_{\rm parent}(\vec{x}_{\rm parent}) \delta t ,      \\
                \vec{p}_{n+1} \rightarrow \vec{v}_{n+1}, \\
                \vec{x}_{n+1} = \vec{x} + \vec{v}_{n+1}\delta t,
            \end{array}\right .
        \end{equation}
        \begin{equation} \label{eq:flip back 2}
            \begin{array}{l}
                \vec{v}_{N,{\rm child, front}} \Rightarrow \vec{v}_{\rm child, front}; - \vec{v}_{N,{\rm child, back}}\Rightarrow \vec{v}_{{\rm child, back}} \\
                \vec{p}_{N,{\rm child, front}} \Rightarrow \vec{p}_{\rm child, front}; - \vec{p}_{N,{\rm child, back}} \Rightarrow \vec{p}_{\rm child, back}
            \end{array}
        \end{equation}
    \item Along a geodesic while updating the metric: the metric and external force are updated according to the child particle location at each integration step and the velocity is updated according to the current momentum and metric. Again the integration is completed with $N \geq 10$ timesteps until $\Delta t = r_{\rm sep}/|v|_{\rm parent}$ for accuracy, with $\delta t = \min{\left(0.1 u_t, 0.1\Delta t, \Delta t_{\rm ext}\right)}$.
    \begin{equation} \label{eq:flip 3}
                \vec{p}_{\rm parent} \Rightarrow \vec{p}_{\rm 0, child, front}; -\vec{p}_{\rm parent} \Rightarrow \vec{p}_{0, \rm child, back}
    \end{equation}
\begin{equation} \label{eq:split}
    \begin{array}{rr}
       \begin{array}{cc}
            \textrm{for each child: front \& behind } \\
            \textrm{loop over } n \textrm{ until } \Delta t \textrm{ is reached} 
        \end{array}
        \left\{
    \begin{array}{l}
            \vec{p}_{n+1} = \vec{p}_{n} + \vec{f}^{\rm ext}_{n}(\vec{x}_{n}) \delta t ,      \\
        \vec{p}_{n+1} \rightarrow \vec{v}_{n+1}, \\
        \vec{x}_{n+1} = \vec{x}_n + \vec{v}_{n+1}\delta t,
    \end{array}\right .&
    \end{array}
\end{equation}
\begin{equation} \label{eq:flip back 3}
            \begin{array}{l}
                \vec{v}_{N,{\rm child, front}} \Rightarrow \vec{v}_{\rm child, front}; - \vec{v}_{N,{\rm child, back}}\Rightarrow \vec{v}_{{\rm child, back}} \\
                \vec{p}_{N,{\rm child, front}} \Rightarrow \vec{p}_{\rm child, front}; - \vec{p}_{N,{\rm child, back}} \Rightarrow \vec{p}_{\rm child, back}
            \end{array}
        \end{equation}
    
    \item Updating the full geodesic: Here we follow the method outlined in Section 6.3 of \citet{Liptai2019a}, updating both particle position and velocity fully with a leapfrog scheme.
\end{enumerate}
Because the above methods constrain the particles to stay along their current geodesic they also naturally enforce the desired conservation properties (energy, angular momentum, etc.) even though the newly refined particles may have different velocities.

\begin{figure}
    \centering
    \includegraphics[width=0.5\linewidth]{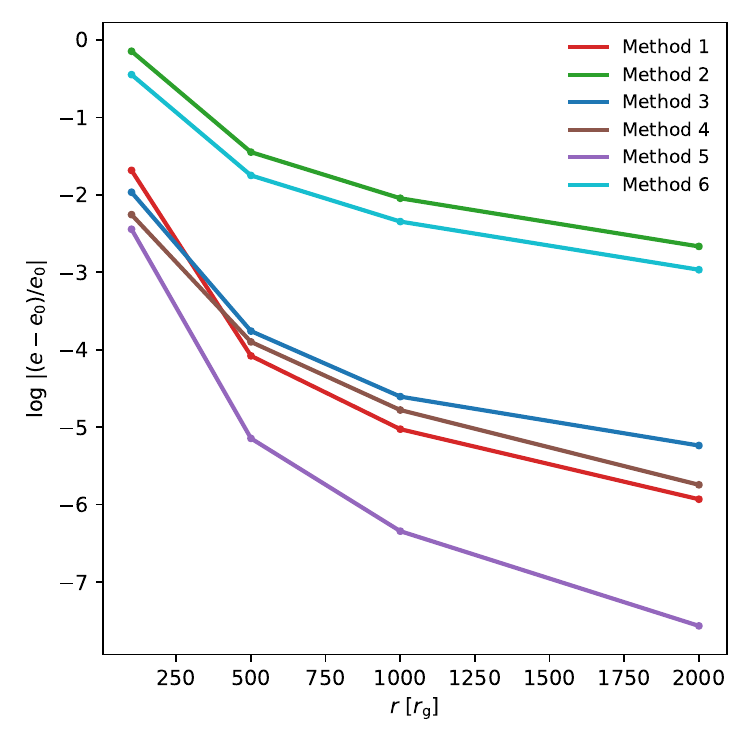}
    \caption{Difference in initial orbital energy and final orbital energy after splitting as a function of distance from the MBH for various splitting methods. The difference in energies is lowest across all radii for the method which splits particles along the geodesic and updates the metric (Method 5) - the method we employ in this work.}    \label{fig:benchmark}
\end{figure}

We quantify the effectiveness of these methods by considering the error in the total energy and the spreading of the stream. Figure~\ref{fig:benchmark} shows the difference between the original and new orbital energy for particles split with the above different methods at different radii. The original APR implementation mostly outperforms the other methods except Method 5, where we update the geodesic and the metric. 
However the original APR implementation artificially widen the stream and put particles onto orbits different to the original, which leads to even more fanning after pericenter. We conclude that the geodesic metric method is superior in conserving the total energy at all radii while faithfully reproducing the width of the debris stream. 
We thus use this method to place newly refined particles and to assign their velocities for all analyses in the main text. Based on Figure~\ref{fig:benchmark}, the original APR implementation could still be the preferred method when the direction of splitting is not critical given it is computationally much cheaper than the new method.

\section{Computing energy contributions at pericenter}
\label{app:energy_appendix}
\begin{figure}
    \centering
    \includegraphics[width=0.49\linewidth]{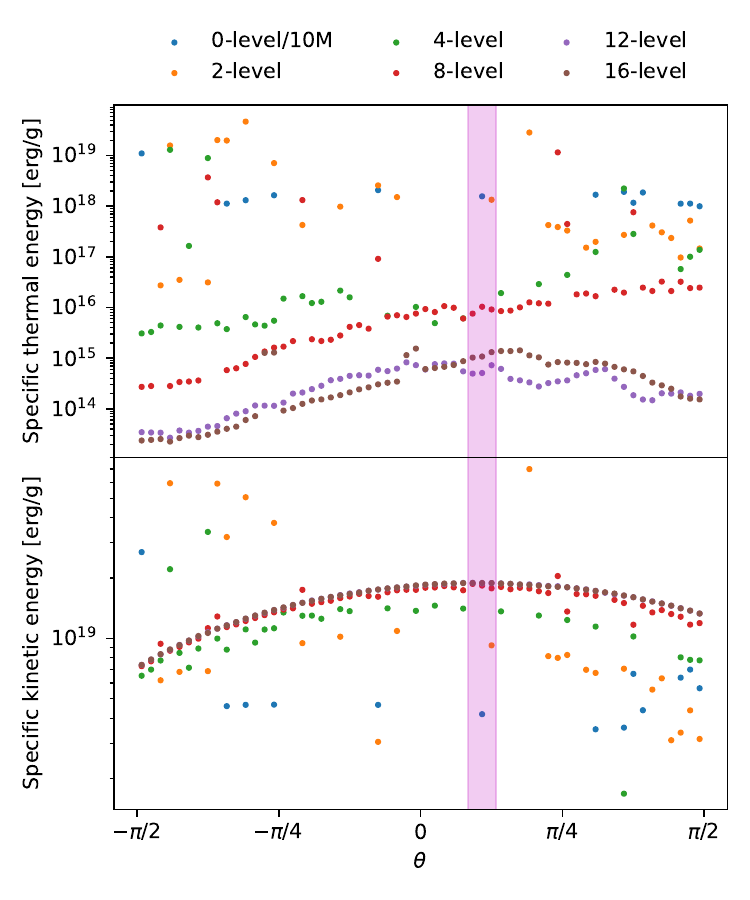}
    \includegraphics[width=0.49\linewidth]{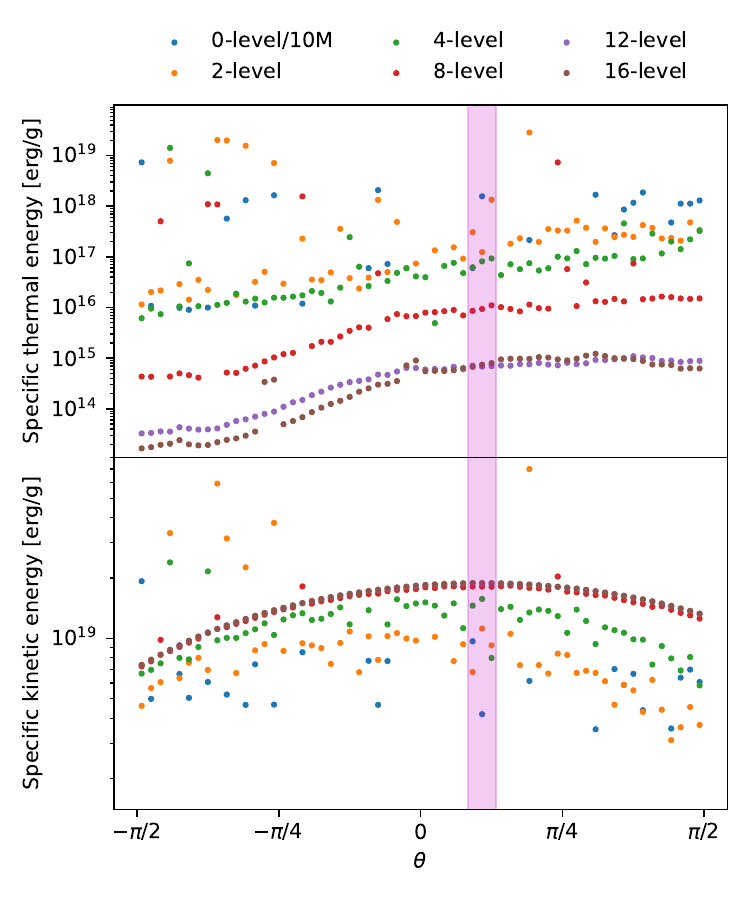}
    \caption{Measuring the specific thermal and kinetic energies as a function of angular coordinate $\theta$ relative to the MBH as the leading edge of the tidal stream passes through pericenter. The pink region represents the estimated pericenter used in Section~\ref{sec:circ eff}. \emph{Left}: one snapshot with very few particles for the 0-, 2- and 4-level simulations.  Particles are only present in 13, 26, 33 beams, respectively, of the 60 beams plotted. \emph{Right}: the average of four adjacent snapshots, showing enough particles to meaningfully analyse in the lower resolution simulations, with 25, 56, 59 beams occupied for 0, 2 and 4 level simulations, respectively. Data is available on Zenodo: 10.5281/zenodo.17225834. 
    }
    \label{fig:angle profile}
\end{figure}

Figure~\ref{fig:angle profile} shows the specific thermal and kinetic energies of the particles as the leading edge of the debris stream passes through pericenter. The location of pericenter is not well defined due to the GR precession, therefore we approximate it as the location with the maximum kinetic energy in the simulation with the highest resolution (16 levels of refinement). The left panel of Figure~\ref{fig:angle profile} shows a single snapshot as the leading edge just passes through pericenter at $t=14.4$ days post disruption, the moment shown in Figure~\ref{fig:split}.  

Since there are too few particles in 0, 2 or 4 level simulations for any meaningful statistical analysis from one snapshot, we choose to stack four snapshots (separated by 4.38 hours) for analysis.  We count all the particles present in each angular beam in the four snapshots. The result is shown in the right-hand panel of Figure~\ref{fig:angle profile}.  We combine three beams, ranging between $\pm 4.5^\circ$ from the approximate pericente, to reconstruct the steam at periapsis (pink regions in Figure~\ref{fig:angle profile}). The number of particles within the three beams of interest when using four (one) snapshots is 3(2), 3(1), 6(0), 66(10), 1036(221), 20403(4225) for the 0, 2, 4, 8, 12 and 16 level simulations, respectively. We then calculate the mean specific thermal energy and kinetic energy and associated standard deviations within the combined three beams over four time snapshots. These results are used to calculate the fraction shown in Figure~\ref{fig:circularisation efficiency} and Section~\ref{sec:circ eff}.


\bibliography{sample631}{}
\bibliographystyle{aasjournal}



\end{document}